\documentclass[11pt]{article}
\usepackage{moriond,epsfig}
\usepackage{amsmath,amssymb,graphicx}
\def\mydate{16 May l 2012}
\def\onehalf{{\frac{1}{2}}}
\def\eff{{\rm eff}}
\def\EM{{\rm EM}}
\def\KK{{\rm KK}}
\def\go{{\rightarrow}}

\def\Journal#1#2#3#4{{#1} {\bf #2}, #3 (#4)}

\def\NPB{{\em Nucl. Phys.} B}
\def\PLB{{\em Phys. Lett.}  B}

\def\PRD{{\em Phys. Rev.} D}

\def\be{\begin{equation}}
\def\ee{\end{equation}}
\def\bea{\begin{eqnarray}}
\def\eea{\end{eqnarray}}

\begin{document}
\vspace*{4cm}
\title{REALISTIC $\mathbf {SO(5) \times U(1)}$ MODEL IN RS SPACE}

\author{YUTAKA HOSOTANI}

\address{Department of Physics,  Osaka University\\
Toyonaka, Osaka 560-0043, Japan}

\maketitle\abstracts{
The gauge bosons and Englert-Brout-Higgs (EBH) boson are unified in the five 
dimensional RS spacetime.
The EBH boson is identified with a part of the fifth dimensional component
of the gauge potential.  In the $SO(5) \times U(1)$ gauge-Higgs unification
the EW symmetry is dynamically broken. The EBH boson, predicted with a mass
around 130 GeV, naturally becomes stable so that it appears as missing energy and
momentum in collider experiments. Collider signatures such as gauge couplings of
quarks and leptons and production of KK $\gamma$ and Z are also discussed.
\hfill (OU-HET 746/2012, \mydate)
}

\vskip -15pt
\centerline{\small \it To appear in the Proceedings of ``Rencontres de Moriond 2012, EW Session''.}

\section{Introduction}
The last particle yet to be discovered in the standard model (SM) of 
strong and electroweak (EW)  interactions is the Englert-Brout-Higgs (EBH) boson.   
Possible signals  for the  EBH boson at the LHC experiments have been 
reported, but more data are necessary for the confirmation.

If the EBH boson is found around 125 GeV, but with non-SM couplings to $W$, $Z$, and
fermions, or if the EBH boson is not seen at LHC, not because it does not exist, but 
because it is stable, then the gauge-Higgs unification  scenario becomes
plausible.   In either case it becomes urgent matter to explore 
the gauge-Higgs unification.

\section{Gauge-Higgs Unification}

We start with gauge theory in higher dimensions where extra-dimensional 
space is not simply connected.  Take a five-dimensional theory.
Zero modes of four-dimensional components of the vector potentials $A_\mu$
contain photon, $W$, and $Z$, whereas zero modes of the extra dimensional
component $A_y$ contain the 4D EBH boson.  Thus the EBH boson 
becomes a part of the gauge bosons, leading to the gauge-Higgs
unification (GHU).\cite{YH1}$^{,\,}$\cite{Davies1}$^{,\,}$\cite{Hatanaka1998} 

When the extra dimensional space is not simply connected, the zero mode
of $A_y$ appears as an Aharonov-Bohm (AB) phase in the extra dimension.
Though its non-vanishing vacuum expectation value (vev) gives vanishing 
field strengths ($F_{\mu y}$), it becomes a physical degree of freedom,
causing dynamical gauge symmetry breaking by the Hosotani mechanism
at the quantum level. 

Symbolically the phase appears as a Wilson line integral along a 
non-contractible loop $C$ in the fifth dimension
\be
e^{i \hat \theta_H (x)} \sim P \exp \bigg\{ ig \int_C dy \, A_y \bigg\}
~~,~~~
\hat \theta (x) = \theta_H + \frac{H(x)}{f_H} ~~.
\label{ABphase1}
\ee
The field $H(x)$ corresponds to the 4D neutral EBH boson to be discovered
at Tevatron/LHC.
The constant part $\theta_H$ corresponds to the ratio of the vev of
the EBH boson to $f_H$.  $\theta_H \not= 0$ induces the EW symmetry
breaking, and at the same time gives masses for quarks, leptons, 
$W$ and $Z$ as in  SM.  A novel, and decisively important, feature is that
$\theta_H$ is a phase. Physical quantities are periodic in $\theta_H$
with a period $2\pi$; 
\be
\theta_H \sim \theta_H + 2 \pi ~.
\label{ABphase2}
\ee
It distinguishes GHU from SM.

\section{$\mathbf {SO(5) \times U(1)}$ Gauge-Higgs Unification in RS space}

Several features must be implemented in a realistic model.  First of all
quark-lepton content must be chiral.  This feature is most easily 
realized if the extra dimensional space has the structure of an orbifold.  
Secondly the model must naturally contain the SM gauge structure
$SU(2)_L \times U(1)_L$.  This is achieved by starting with the gauge
group $SO(5) \times U(1)_X$ which incorporates the custodial 
symmetry.   Thirdly the EW symmetry must be dynamically broken,
which is achieved, with minimal fermion content, in the Randall-Sundrum (RS)
warped spacetime;
\be
ds^2 = e^{- 2 \sigma (y)} dx^\mu dx_\mu + dy^2
\label{RS1}
\ee
where $\sigma (y) = k y$ for $0 \le y \le L$ and 
 $\sigma (y) = \sigma (-y)= \sigma (y + 2L)$.  
The warp factor is given by $z_L = e^{kL}$. 
The bulk spacetime $0 < y < L$ is AdS spacetime with the curvature $- 6 k^2$,
which is sandwiched by the Planck brane at $y=0$ and the TeV brane at
$y=L$.  It has topology of $M^4 \times (S^1/Z_2)$.

The bulk part of the action  consists of the $SO(5)$ and $U(1)_X$ gauge 
fields $A_M$ and $B_M$ with gauge couplings $g_A$ and $g_B$, 
and bulk fermions.  The bulk fermions $\Psi_a$ are introduced in vector 
representation of $SO(5)$.  In each generation two multiplets in the quark 
sector and two multiplets in the lepton sector are introduced. 
They satisfy the orbifold boundary conditions
\bea
&&\hskip -1.cm
\begin{pmatrix} A_\mu \cr A_y \end{pmatrix}(x, y_j -y) =
P_j \begin{pmatrix} A_\mu \cr - A_y \end{pmatrix}(x, y_j +y)  P_j^{-1} ~, \cr
\noalign{\kern 10pt}
&&\hskip -1.cm
(B_\mu, B_y) (x, y_j -y) = (B_\mu,-  B_y) (x, y_j +y) ~, \cr
\noalign{\kern 10pt}
&&\hskip -1.cm
\Psi_a (x, y_j -y) = P_j \gamma^5 \Psi_a (x, y_j +y) ~,
\label{BC1}
\eea
where $(y_0. y_1) = (0, L)$ and $P_j = P_j^\dagger = P_j^{-1}$.  
In particular we take $P_j = {\rm diag} \,  (-1, -1, -1, -1, +1)$,
which reduces the symmetry to $SO(4) \times U(1)_X$.
In addition brane fermions and brane scalar are introduced on the
Planck brane.  The brane scalar, which is $(0, \onehalf)$ representation of
$SO(4) \simeq SU(2)_L' \times SU(2)_R'$, spontaneously breaks the symmetry
$SU(2)_R' \times U(1)_X$ to $U(1)_Y'$ and makes all exotic fermions heavy. 
The resultant symmetry is $SU(2)_L' \times U(1)_Y'$.   
The $SO(4) \times U(1)_X$ chiral anomalies are cancelled with the brane 
fermions.   After the symmetry breaking by $\theta_H$,
the low energy spectrum is the same as in SM.\cite{ACP}$^{-}$\cite{HTU1}

\section{Dynamical EW Symmetry Breaking by the Hosotani Mechanism}

One of the nicest features in this model is that the EW symmetry is dynamically
broken to the electromagnetic $U(1)_{EM}$ by the Hosotani 
mechanism.\cite{HOOS}$^{,\,}$\cite{HTU1}
For the dynamical EW symmetry breaking it is crucial that (i)
the  multiplet containing a top quark is in the vector representation of $SO(5)$,
and (ii) the spacetime is Randall-Sundrum warped space, but is not  flat.

The effective potential $V_\eff$ for $\theta_H$ at the one-loop level is depicted 
in fig.\ 1.  In the pure gauge theory the symmetry remains unbroken.  In the
presence of the top quark, whose mass is  larger than $m_W$, $V_\eff$ is 
minimized at $\theta_H=\pm \onehalf \pi$ so that the EW symmetry breaks down to 
$U(1)_\EM$.

\begin{figure}[t,b]
\centering  \leavevmode
\includegraphics[height=4.3cm]{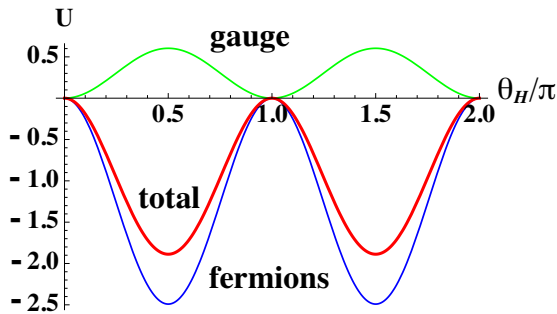}
\caption{The effective potential  $V_\eff (\theta_H)$. The plot is for  
$U (\theta_H/\pi) = (4\pi)^2 (kz_L^{-1})^{-4} \, V_\eff$  at $z_L = 10^{15}$.  
Green, blue, and red curves represent $V_\eff^{\rm gauge}$, $V_\eff^{\rm fermion}$,
and $V_\eff$, respectively. 
The global minima are located at 
$\theta_H = \onehalf \pi $ and $\frac{3}{2}\pi$, where  the EW symmetry  
dynamically breaks down to $U(1)_{\EM}$.}
\label{effV-total-fig}
\end{figure}

The EBH boson mass $m_H$ is given by 
$m_H^2 = f_H^{-2} (d^2 V_\eff/d\theta_H^2)$ at the minimum.
It is found that $m_H = 135 \, (72) \,$GeV for $z_L = 10^{15} \, (10^5)$.
This does not contradict with the current experimental data,
since the EBH boson becomes stable as is seen below.

\section{Effective Low-Energy Interactions}

The effective Lagrangian at low energies among the EBH boson, $W$, $Z$,
quarks and leptons is approximately given 
by \cite{Sakamura2007}$^{,\,}$\cite{HK2009}$^{,\,}$\cite{Giudice}  
\be
{\cal L}_\eff \sim - \Big( \onehalf g f_H \sin \hat \theta_H \Big)^2 
\bigg\{ W_\mu^\dagger W^\mu + \frac{1}{2 \cos^2 \theta_W} Z_\mu Z^\mu \bigg\}
- y_f f_H \sin \hat \theta_H \overline{\psi}_f \psi_f  
\label{effLag1}
\ee
where $\hat \theta_H $ is given by (\ref{ABphase1}) and 
$\onehalf g f_H = m_{\KK} / \pi \sqrt{kL}$.  The expression is valid to good accuracy
for  large $z_L$.  The Kaluza-Klein mass is given by $m_{\KK} = \pi k z_L^{-1}$,
which turns out to be around 1.4$\,$GeV for $z_L = 10^{15}$.  
In SM one has $v + H$ in place of
$f_H \sin \hat \theta_H$.   The nature of $\theta_H$ as an AB phase forces
the appearance of periodic, non-linear mass functions, a distinguishing feature of
the gauge-Higgs unification.  
$f_H = 246\,$GeV for $\theta_H = \pm \onehalf \pi$.

The masses are given by $m_W = \onehalf g f_H | \sin \theta_H |$, 
$m_Z = m_W/\cos \theta_W$, and $m_f = y_f | \sin \theta_H |$. 
The couplings of the EBH boson to $W$, $Z$, quarks and leptons are obtained
by expanding the expression (\ref{effLag1}) in a Taylor series in $H$.  
The linear couplings  are found to be
\be
WWZ, ZZH, {\rm Yukawa ~ couplings}
= ({\rm SM ~ values}) \times \cos \theta_H ~.
\label{coupling1}
\ee
They are suppressed, compared with the SM values, by a universal factor $\cos \theta_H$.  
This is a specific character of the gauge-Higgs unification.\cite{Lim2012}
In particular, the linear couplings vanish at $\theta_H = \onehalf \pi$.
We stress that $\theta_H \not= 0$ gives masses to $W$, $Z$ and fermions
as in SM,  but gives vanishing linear couplings at $\theta_H = \onehalf \pi$.

\section{H Parity and Stable EBH Bosons}

The fact that the $WWH$, $ZZH$ and Yukawa couplings vanish
at $\theta_H = \onehalf \pi$ is a consequence of the symmetry.
There emerges $H$-parity,  $P_H$,  at $\theta_H = \onehalf$.
Among low-energy particles the EBH boson is odd under $P_H$,
while all other SM particles are even.  It immediately follows that the EBH boson
becomes absolutely stable.\cite{HKT}$^{,\,}$\cite{HTU1}  

The proof for the existence of the $H$-parity proceeds as follows.
First the action of the model is invariant under the mirror reflection
in the fifth dimension;  $(x^\mu, y ) \go (x^\mu, -y)$, 
$(A_\mu, A_y) \go$ $ (A_\mu, - A_y)$, and $\Psi \go \pm \gamma^5 \Psi$.
Under the reflection $\hat \theta_H \go - \hat \theta_H$, while wave functions
of all other SM particles remain invariant.
Secondly, there arises the enhanced large gauge symmetry when all bulk
fermions belong to the vector representation of $SO(5)$.  The periodicity
in $\theta_H$ in physical quantities is halved to $\pi$; 
$\theta_H + \pi \sim \theta_H$.  If there were a fermion in the spinor
representation of $SO(5)$, the periodicity would remain as the original
$2 \pi$.  Thirdly the effective potential $V_\eff (\theta_H)$ is minimized
at $\theta_H = \onehalf \pi$ thanks to the presence of the top quark.
Around $\theta_H = \onehalf \pi$ we have, for physical quantities, 
equivalence relations
\be
\frac{\pi}{2} + \frac{H}{f_H}  ~ \Leftrightarrow  ~
- \frac{\pi}{2} - \frac{H}{f_H}  ~ \Leftrightarrow ~
\frac{\pi}{2} - \frac{H}{f_H}  ~.
\label{equivalence}
\ee
The theory is invariant under $P_H$ to all orders in perturbation theory.
The symmetry around $\onehalf \pi$ has been seen, for instance, 
in $V_\eff(\theta_H)$ depicted in fig.\ 1.

Another proof for the $H$-parity has been provided, by noticing the 
invariance of the $SO(5)$ algebra under the interchange of
$SU(2)_L'$ and $SU(2)_R'$ and flip $T^{\hat 4} \go - T^{\hat 4}$.
At $\theta_H = \onehalf \pi$  the brane fields couple to only bulk fields 
which are even under this operation.   This symmetry suppresses radiative corrections
to the $T$ parameter and $Zb \bar b$ coupling as noticed by Agashe et al.\cite{Agashe2006}

With the $H$-parity all $H^n$-couplings ($n$: an odd integer) 
to other SM particles are forbidden.  
The LEP2 constraint for the EBH boson mass ($m_H \ge 114\,$GeV) 
is also evaded as the $ZZH$ coupling vanishes.

\section{Collider Signatures}

The phenomenology at $\theta_H = \onehalf \pi$ is extremely 
interesting.\cite{Cheung2010a}$^{-}$\cite{Contino2011}
The $H$-parity forbids production of a single EBH boson.  EBH bosons 
are produced in pairs at collider experiments.  The production rate is normal.
The $WWHH$, $ZZHH$ couplings are $-1$ times the SM couplings.
The EBH boson becomes stable.  It implies that produced EBH bosons
do not decay so that EBH bosons appear as missing energies and momenta
in collider events. 
As a pair of EBH bosons, two stable particles,  are produced, 
confirming them at Tevatron/LHC/ILC becomes very difficult. 
There are large background events containing neutrinos with the same topology.
If polarized right-handed electron and left-handed positron beams can be prepared 
at ILC, then identification of stable EBH bosons becomes feasible by suppressing
neutrino backgrounds.

The $\chi^2$ values for the forward-backward asymmetry $A_{FB}$ on 
the $Z$ resonance in the $e^+ e^-$ annihilation and for the branching fractions of the $Z$
decay are tabulated in Table 1.  Although the gauge-Higgs unification scenario gives 
good agreement for $A_{FB}$ in a wide range of $z_L$, the branching fractions
of the $Z$ decay are reproduced only for large $z_L \ge 10^{15}$.

\begin{table}[htb]
\begin{center}
\caption{$\chi^2$ fit for $A_{FB}$ and  $Z$ decay 
fractions.
The values of $m_\KK$, $m_H$ and $m_W^{\rm tree}$
 are also listed.
}
\label{chi2fit}
\vskip 10pt
\begin{tabular}{|c||c|c|c|c||c|}
\hline
&\# of data & $z_L=10^{15}$ & $10^{10}$ & $10^5$ &SM \\
\hline
$\sin^2 \theta_W$ &&0.2309 & 0.2303 & 0.2284 & 0.2312 \\
\hline
$\chi^2$ [$A_{FB}$] & 6 
& 
6.3
& 
6.4
&
7.1
&
10.8
\\
$\chi^2$ [$Z$ decay fractions] & 8
& 
16.5
& 
37.7
& 
184.5
& 
13.6
\\
\hline 
Sum of two $\chi^2$ & 14
& 
22.8
&
44.1
& 
191.6
& 
24.5
\\
\hline 
$m_\KK$ (GeV) &  & 1466 & 1193 & 836 &\\
$m_H$ (GeV) &  &135 &108 &72 &\\
$m_W^{\rm tree}$ (GeV) &  &79.84 &79.80 &79.71 &79.95 \\
\hline
\end{tabular}
\end{center}
\end{table}

The signatures of the extra dimension itself are obtained by observing
KK excited states of various particles.  Relatively clear signals can be found
for KK $Z^{(1)}$ and $\gamma^{(1)}$, which subsequently decay into
$e^+ e^-$ or $\mu^+ \mu^-$.  The masses and total decay widths of
$Z^{(1)}$ and $\gamma^{(1)}$ are tabulated in Table 2.
Unlike other conventional models the current gauge-Higgs unification model
predicts large production rates and decay widths for KK gauge bosons.
This is because right-handed quarks and leptons have large couplings
to the KK gauge bosons.    
KK $Z^{(1)}$ corresponds to what is referred to as $Z'$ in the analyses of
Tevatron and LHC data.  So far no signal of $Z'$ has been found.
This may indicate the necessity for improving the current gauge-Higgs unification model.

\begin{table}[htb]
\begin{center}
\caption{Masses and widths of KK $Z^{(1)}$ and $\gamma^{(1)}$.}
\label{KKgauge}
\vskip 5pt
\begin{tabular}{|c|c|c|}
\hline 
\multicolumn{3}{|c|}{$Z^{(1)}$}
\\
\hline
$z_L$ &  $10^5$  & $10^{15}$ 
\\
\hline
$m$ (GeV)&  653  & 1130 
\\
\hline
$\Gamma$ (GeV) &  104  & 422 
\\
\hline
\end{tabular}
\quad
\begin{tabular}{|c|c|c|}
\hline 
\multicolumn{3}{|c|}{$\gamma^{(1)}$}
\\
\hline
$z_L$ &  $10^5$  & $10^{15}$ 
\\
\hline
$m$ (GeV)&  678  & 1144 
\\
\hline
$\Gamma$ (GeV) &  446  & 1959 
\\
\hline
\end{tabular}
\end{center}
\end{table}

\section{Stable EBH Bosons as Dark Matter v.s. Supersymmetry}

EBH bosons become stable at $\theta_H = \onehalf \pi$. 
They are copiously produced in the early universe.  As the universe expands
and the annihilation rate of EBH bosons falls, the annihilation processes get frozen and
the remnant EBH bosons become dark matter.\cite{HKT}  The annihilation
couplings are determined from the effective interaction (\ref{effLag1}).
The present mass density of cold dark matter has been determined by WMAP collaboration as
$\Omega_{\rm CDM} h^2 = 0.1131 \pm 0.0034$. 

Suppose that the EBH mass is sufficiently smaller than  $m_W$. 
In this case the dominant annihilation process is $HH \go b \bar b$, and
the abundance  turns out much larger than the WMAP value. 
If the EBH boson is heavier than $W$,
$HH \go W^+W^-$ dominates, and the relic abundance turns out 
much smaller than the WMAP value. 
The cold dark matter abundance observed by WMAP
is reproduced with $m_H = 70 \sim 75 \,$GeV. 

This is a very attractive scenario; the EBH boson responsible for the EW symmetry
breaking constitutes the dark matter of the universe.  However, the EBH boson
mass $m_H = 70 \sim 75 \,$GeV  is realized in the current model with the warp
factor $z_L \sim 10^5$, which conflicts with the precision measurements of the 
gauge couplings at low energies and collider data at high energies as  seen above.   

Of course nothing is wrong with the scenario of the gauge-Higgs unification
with $z_L \ge 10^{15}$ in which the dark matter is accounted for by other
particles.  More ambitiously we  ask if it is possible to have the gauge-Higgs
unification in which, without conflicting with the collider data, stable
EBH bosons account for the cold dark matter abundance observed by WMAP.

We argue that it is possible, provided supersymmery (SUSY) exists.\cite{HH2011} 
If SUSY is exact and unbroken, then the EBH boson remains massless as
boson and fermion contributions to $V_\eff (\theta_H)$ cancell each other.
SUSY is softly or dynamically broken so that the cancellation becomes
incomplete, leading to a non-vanishing $m_H$.

Conversely one can determine SUSY breaking scales such that 
the EBH boson acquires a mass around $70 \sim 75\,$GeV with a given
warp factor, say, $z_L = 10^{15}$.  At the one loop level only the spectra of
SUSY partners of $W$, $Z$,  top quark  and their
KK towers are relevant.  It is found that the scenario of light neutralinos ($< 100\,$GeV),
heavy gluinos ($> 1\,$TeV), and the stop with $m_{\rm stop} = 300 \sim
320\,$GeV yield the desired mass $m_H = 70 \sim 75\,$GeV at $z_L = 10^{15}$.
Finding a stop may give a hint for extra dimensions.

\section{Summary}

We have seen that the minimal $SO(5) \times U(1)$ gauge-Higgs unification
model in RS leads to astonishing prediction that the EBH boson is stable.
The EBH boson is identified with a part of the gauge potentials in the extra dimension.
It appears as four-dimensional fluctuations of the AB phase in the extra dimension.
The nature as an AB phase gives significant deviation from SM, which can be
tested at colliders.  We may be about to see the extra dimension at LHC.

\section*{Acknowledgments}
This work was supported in part 
by  scientific grants from the Ministry of Education and Science, 
Grants No.\ 20244028, No.\ 23104009 and  No.\ 21244036.

\def\AP{{\em Ann.\ Phys.\ (N.Y.)} }
\def\MPLA{{\em Mod.\ Phys.\ Lett.} A}
\def\PTP{{\em Prog.\ Theoret.\ Phys.}}
\def\ibid{{\em ibid.} }
\def\JHEP{{\em JHEP} }

\end{document}